# A Convex Cauchy-Schwarz Divergence Measure for Blind Source Separation

Zaid Albataineh[1] and Fathi M. Salem[2]

*Abstract*—We propose a new class of divergence measures for Independent Component Analysis (ICA) for the *demixing* of multiple source mixtures. We call it the Convex Cauchy-Schwarz Divergence (CCS-DIV), and it is formed by integrating convex functions into the Cauchy-Schwarz inequality. The new measure is symmetric and the degree of its curvature with respect to the joint-distribution can be tuned by a (convexity) parameter. The CCS-DIV is able to speed-up the search process in the parameter space and produces improved demixing performance. An algorithm, generated from the proposed divergence, is developed which is employing the non-parametric Parzen window-based distribution. Simulation evidence is presented to verify and quantify its superior performance in comparison to state-of-the-art approaches.

**Index Terms**—Independent Component Analysis (ICA), Cauchy-Schwarz inequality, non-parametric Independent Component Analysis (ICA), Parzen window-based distribution.

## I. INTRODUCTION

Blind Signal Separation (BSS) is one of the most challenging and emerging areas in signal processing. BSS has a solid theoretical foundation and numerous potential applications. BSS remains a very important and challenging area of research and development in many domains, e.g. biomedical engineering, image processing, communication system, speech enhancement, remote sensing, etc. BSS techniques do not assume full a priori knowledge about the mixing environment, source signals, etc. and do not require training samples. Independent Component Analysis (ICA) is considered a key approach in BSS and unsupervised learning algorithms [1], [2].

ICA specializes to Principal Component Analysis (PCA) and Factor Analysis (FA) in multivariate analysis and data mining, corresponding to second order methods in which the components are in the form of a Gaussian distribution [6 - 9], [1], [2]. However, ICA is a statistical technique that exploits higher order statistics (HOS), where the goal is to represent a set of random variables as a linear transformation of statistically independent components.

We provide a brief overview relevant to this letter. The metrics of cumulants, likelihood function, negentropy, kurtosis, and mutual information have been developed to obtain a demixing matrix in different adaptations of ICA-based algorithms [1]. Recently, Zarzoso and Comon [7] proposed Robust Independent Component Analysis (R-ICA). He used a truncated polynomial expansion rather than the output marginal probability density functions to simplify the estimation processes. In [10 – 12], the authors have presented ICA using mutual information. They constructed a formulation by minimizing the difference between the joint entropy and the marginal entropies of signals.

The so-called convex ICA [13] is established by incorporating a convex function into a Jenson's inequality-based divergence measure. Xu et al. [14] used the approximation of Kullback–Leibler (KL) divergence based on the Cauchy–Schwartz inequality. Boscolo et al. [15] established nonparametric ICA by minimizing the mutual information contrast function and by using the Parzen window distribution.

A new contrast function based on a nonparametric distribution was developed by Chien and Chen [16], [17] to construct an ICA-based algorithm. They used the cumulative distribution function (CDF) to obtain a uniform distribution from the observation data. Moreover, Matsuyama et al. [18] proposed the alpha divergence approach. Also, the f-divergence was proposed by Csiszár et al. [3], [19].

In addition, the maximum-likelihood (ML) criterion [21] is another tool for BSS algorithms [21]–[23]. It is used to estimate the demixing matrix by maximizing the likelihood of the observed data. However, the ML estimator needs to know (or estimate) all the source distributions. Recently, in terms of divergence measure, Fujisawa et al. [24] have proposed a very robust similarity measure to outliers and they called it the Gamma divergence. In addition, the Beta divergence was proposed in [25] and investigated by others in [3].

Xu et al. [5] proposed the quadratic divergence based on the Cauchy-Schwartz inequality, namely, Cauchy-Schwartz divergence (CS-DIV). CS-DIV is used to implement the ICA procedure, but it lacks the optimality and the stability in terms of performance since the CS-DIV is not a convex divergence.

While there are numerous measures, performance in terms of the quality of the estimated source signals still in need of improvements. Thus, the present work focuses on enhancing the performance in terms of the quality of the estimated demixed signals. To that end, we develop a new class of divergence measures for ICA algorithms based on the conjunction of a convex function into a Cauchy–Schwarz inequality-based divergence measure. This symmetric measure has a wide range of effective curvatures since its curvature is controlled by a convexity parameter. With this convexity, unlike CS-DIV, the proposed measure is more likely to attain an optimal solution and speed up the convergence in the separation process. As a result, the proposed divergence results in better performance than other methods, especially CS-DIV. Moreover, it is considered to be an effective alternative measure to Shannon's mutual information measure. The convex Cauchy–Schwarz divergence ICA (CCS–ICA) uses the Parzen window density to distinguish the non-Gaussian structure of source densities. The CCS-ICA has succeeded in solving the BSS of speech and Music signals with and without additive noise and it has shown a better performance than other ICA-based methods. *Finally, it is important to highlight that while the divergence measure is convex with respect to the joint probability density, it is only locally convex with respect to the filtering parameters. It is well-known that the BSS problem has a (scaling and) permutation ambiguity and thus there are multiple solutions.*

[1] Electronics Engineering Department, Hijjawi Faculty for Engineering Technology, Yarmouk University, Irbid, Jordan
Email: albatain@msu.edu

[2] Circuits, Systems, And Neural Networks (CSANN) Laboratory
Department of Electrical and Computer Engineering, Michigan State University, East Lansing, Michigan 48824-1226, U.S.A.
Emails: salemf@msu.edu

The letter is organized as follows. Section II proposes the new convex Cauchy–Schwarz divergence measure. Section III presents the CCS–ICA method. The comparative simulation results and conclusions are given in Section IV and Section V, respectively.

## II. A Brief Description of Previous Divergence Measures

Divergence, or the related (dis)similarity, measures play an important role in the areas of neural computation, pattern recognition, learning, estimation, inference, and optimization [3]. In general, they measure a quasi-distance or directed difference between two probability distributions which can also be expressed for unconstrained arrays and patterns. Divergence measures are commonly used to find a distance between two n-dimensional probability distributions, say $p = (p_1, p_2, \ldots p_n)$ and $q = (q_1, q_2, \ldots q_n)$. Such a divergence measure is a fundamental key factor in measuring the dependency among observed variables and generating the corresponding ICA-based procedures.

A *metric* is the distance between two pdfs if the following conditions hold: $(i)\ D(p||q) = \sum_{i=1}^{n} d(p_i, q_i) \geq 0$ with equality if and only if $p = q$, $(ii)\ D(p||q) = D(q||p)$ and $(iii)$ the triangular inequality, i.e., $D(p||q) \leq D(p||z) + D(z||q)$, for another distribution $z$. Distances which are not a metric are referred to as divergences [3].

This paper considers on distance-type divergence measures that are separable, thus, satisfying the condition $D(p||q) = \sum_{i=1}^{n} d(p_i, q_i) \geq 0$ with equality holds if and only if $p = q$. But they are not necessarily symmetric as in condition (ii) above, nor do necessarily satisfy the triangular inequality as in condition (iii) above.

Usually, the vector $p$ corresponds to the observed data and the vector $q$ is the estimated or expected data that are subject to constraints imposed on the assumed models. For the BSS (ICA and NMF) problems, $p$ corresponds to the observed sample data matrix $X$ and $q$ corresponds to the estimated sample matrix $Y = WX$. Information divergence is a measure between two probability curves. In other words, the distance-type measures under consideration are not necessarily a metric on the space $P$ of all probability distributions [3].

Next, we propose a novel divergence measures with one-dimensional probability curves.

### A. New Divergence Measure

While there exist a wide range of measures, performance especially in audio and speech applications still requires improvements. The quality of an improved measure should provide geometric properties for a contrast function in anticipation of a dynamic (e.g., gradient) search in a parameter space of the demixing matrices. The motivation here is to introduce a simple measure and incorporate controllable convexity in order to control convergence to an optimal solution. To improve the performance of the divergence measure and to speed up convergence, we have conjugated a convex function into the Cauchy–Schwarz inequality. In this context, one takes advantage of the convexity's parameter, say alpha, to control the convexity of the divergence function and to speed up the convergence in the corresponding ICA and NMF algorithms. For instance, incorporating the joint distribution $(P_J = p(z_1, z_2))$ and the marginal distributions $(Q_M = p(z_1)p(z_2))$ into the convex function, say, $f(.)$ and conjugating them to the Cauchy–Schwarz inequality yields

$$\left|\langle f(P_J), f(Q_M)\rangle\right|^2 \leq \langle f(P_J), f(P_J)\rangle \cdot \langle f(Q_M), f(Q_M)\rangle$$

$$\left|\langle f(p(z_1, z_2)), f(p(z_1)p(z_2))\rangle\right|^2 \leq \langle f(p(z_1, z_2)), f(p(z_1, z_2))\rangle \\ \cdot \langle f(p(z_1)p(z_2)), f(p(z_1)p(z_2))\rangle$$

(1)

where $\langle \cdot, \cdot \rangle$ is an inner product; $f(.)$ is a convex function, e.g.,

$$f(t) = \frac{4}{1-\alpha^2}\left[\frac{1-\alpha}{2} + \frac{1+\alpha}{2}t - t^{\frac{1+\alpha}{2}}\right] \text{for } t \geq 0 \quad (2)$$

Now, based on the Cauchy–Schwartz inequality a new symmetric divergence measure is proposed, namely:

$$D_{CCS}(P_J, Q_M, \alpha) = \log \frac{\iint f^2(P_J) dz_1 dz_2 \cdot \iint f^2(Q_M) dz_1 dz_2}{\left[\iint f(P_J) \cdot f(Q_M) dz_1 dz_2\right]^2}$$
$$= \log \frac{\iint f^2(p(z_1, z_2)) dz_1 dz_2 \cdot \iint f^2(p(z_1) \cdot p(z_2)) dz_1 dz_2}{\left[\iint f(p(z_1, z_2)) \cdot f(p(z_1)p(z_2)) dz_1 dz_2\right]^2}$$

(3)

where, as usual, $D_{CCS}(P_J, Q_M, \alpha) \geq 0$ and equality holds if and only if $p(z_1) = p(z_2)$. This divergence function is then used to develop the corresponding ICA and NMF algorithms. We note that the joint distribution and the product of the marginal densities in $D_{CCS}(P_J, Q_M, \alpha)$ is symmetric. This symmetrical property does not hold for the KL-DIV, α-DIV, and f-DIV. We anticipate that this symmetry would be desirable in the geometric structure of the search space to exhibit similar dynamic trajectories towards a minimum. Additionally, the CCS-DIV is tunable by the convexity parameter α. In contrast to the C-DIV [13] and the α-DIV [18], the range of the convexity parameter α is extendable. However, based on l'Hôpital's rule, one can derive the realization of CCS-DIV for the case of $\alpha = 1$ and $\alpha = -1$ by finding the derivatives, with respect to $\alpha$, of the numerator and denominator for each parts of $\mathrm{D_{CCS}}(P_J, Q_M, \alpha)$. Thus, the CCS-DIV with $\alpha = 1$ and $\alpha = -1$ are respectively given in (4) and (5).

### B. Link to other Divergences:

This CCS-DIV distinguishes itself from previous divergences in the literature by incorporating the convex function into (not merely a function of) the Cauchy Schwarz inequality. The paper develops a framework for generating a family of dependency measure based on conjugating a convex function into the Cauchy Schwarz inequality. Such convexity is anticipated (as is evidenced by experiments) to reduce local minimum near the solution and enhance searching then on-linear surface of the contrast function. The motivation behind this divergence is to render the CS-DIV to be convex similar to the f-DIV. For this work, we shall focus on one convex function f(t) as in (2), and its corresponding CCS-DIVs in (3), (4) and (5). It can be seen that the CCS-DIV, for the $\alpha = 1$ and $\alpha = -1$ cases, is implicitly based on Shannon entropy (KL divergence) and Renyi's quadratic entropy, respectively. Also, it is to show that the CCS_DIVs for the $\alpha = 1$ and $\alpha = -1$ cases are convex functions in contrast to the CS-DIV. (See Fig. 2 and sub-section E in the next page.)

### C. Geometrical Interpretation of the Proposed Divergence for $\alpha = 1$ and $\alpha = -1$.

For compactness, let's define the following terms:

$$V_J = \iint (p(z_1,z_2))^2 dz_1 dz_2$$

$$V_M = \iint (p(z_1)p(z_2))^2 dz_1 dz_2$$

$$V_c = \iint p(z_1,z_2)p(z_1)p(z_2) dz_1 dz_2$$

$$V_{JJ} = \begin{cases} \iint \left\{ \begin{pmatrix} p(z_1,z_2) \cdot \log(p(z_1,z_2)) \\ -p(z_1,z_2) + 1 \end{pmatrix}^2 \right\} dz_1 dz_2 & \alpha = 1 \\ \iint \left\{ \begin{pmatrix} \log(p(z_1,z_2)) \\ -p(z_1,z_2) + 1 \end{pmatrix}^2 \right\} dz_1 dz_2 & \alpha = -1 \end{cases}$$

$$V_{MM} = \begin{cases} \iint \left\{ \begin{pmatrix} p(z_1)p(z_2) \cdot \log(p(z_1)p(z_2)) \\ -p(z_1)p(z_2) + 1 \end{pmatrix}^2 \right\} dz_1 dz_2 & \alpha = 1 \\ \iint \left\{ \begin{pmatrix} \log(p(z_1)p(z_2)) \\ -p(z_1)p(z_2) + 1 \end{pmatrix}^2 \right\} dz_1 dz_2 & \alpha = -1 \end{cases}$$

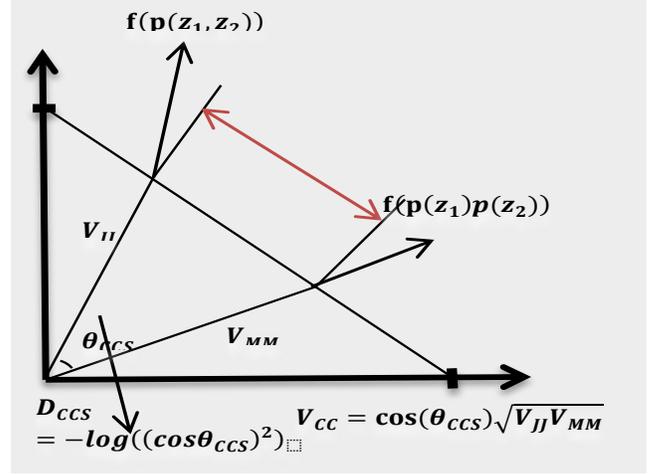

Fig.1: Illustration of the Geometrical Interpretation of the proposed Divergence

$$V_{CC} = \begin{cases} \iint \left\{ \begin{pmatrix} p(z_1,z_2) \cdot \log(p(z_1,z_2)) \\ -p(z_1,z_2) + 1 \end{pmatrix} \cdot \\ \begin{pmatrix} p(z_1)p(z_2) \cdot \log(p(z_1)p(z_2)) \\ -p(z_1)p(z_2) + 1 \end{pmatrix} \right\} dz_1 dz_2 & \alpha = 1 \\ \iint \left\{ \begin{pmatrix} \log(p(z_1,z_2)) \\ -p(z_1,z_2) + 1 \end{pmatrix} \cdot \\ \begin{pmatrix} \log(p(z_1)p(z_2)) \\ -p(z_1)p(z_2) + 1 \end{pmatrix} \right\} dz_1 dz_2 & \alpha = -1 \end{cases}$$

With these terms, one can express the CCS-DIV and the CS-DIV as

$$D_{CCS} = \log(V_{JJ}) + \log(V_{MM}) - 2\log(V_{CC}) \quad (6)$$

$$D_{CS} = \log(V_J) + \log(V_M) - 2\log(V_C) \quad (7)$$

In Fig. 1, we illustrate the geometrical interpretation of the proposed divergence (CCS-DIV), which is equivalent to the Cauchy Schwarz Divergence (CS-DIV). Geometrically, we can show that the angle between the Joint pdfs and marginal pdfs in the CCS-DIV is given as following:

$$\mathrm{D_{CCS}}(P_J, Q_M, 1) = \log \frac{\left( \iint \left\{ (p(z_1,z_2) \cdot \log(p(z_1,z_2)) - p(z_1,z_2) + 1)^2 \right\} dz_1 dz_2 \right) \cdot \left( \iint \left\{ (p(z_1) \cdot p(z_2) \cdot \log(p(z_1) \cdot p(z_2)) - p(z_1) \cdot p(z_2) + 1)^2 \right\} dz_1 dz_2 \right)}{[\iint \{(p(z_1,z_2) \cdot \log(p(z_1,z_2)) - p(z_1,z_2) + 1) \cdot (p(z_1) \cdot p(z_2) \cdot \log(p(z_1) \cdot p(z_2)) - p(z_1) \cdot p(z_2) + 1)\} dz_1 dz_2]^2}$$

(4)

$$\mathrm{D_{CCS}}(P_J, Q_M, -1) = \log \frac{\left( \iint \left\{ (\log(p(z_1,z_2)) - p(z_1,z_2) + 1)^2 \right\} dz_1 dz_2 \right) \cdot \left( \iint \left\{ (\log(p(z_1) \cdot p(z_2)) - p(z_1) \cdot p(z_2) + 1)^2 \right\} dz_1 dz_2 \right)}{[\iint \{(\log(p(z_1,z_2)) - p(z_1,z_2) + 1) \cdot (\log(p(z_1) \cdot p(z_2)) - p(z_1) \cdot p(z_2) + 1)\} dz_1 dz_2]^2}$$

(5)

$$\theta_{CCS} = \mathrm{acos}\left(\frac{V_{CC}}{\sqrt{V_{JJ}V_{MM}}}\right) \equiv \theta_{CS} = \mathrm{acos}\left(\frac{V_C}{\sqrt{V_J V_M}}\right) \quad (8)$$

where $acos$ denotes the cosine inverse. As a matter of fact, the convex function $f$ renders the CS-DIV a convex contrast function for the $\alpha = 1$ and $\alpha = -1$ cases. Moreover, in practice, it provides the proposed measure an advantage over the CS-DIV in terms of potential speed and accuracy, see fig. 2.

*D. Evaluation of Divergence Measures*

In this section, the relations among the KL-DIV, E-DIV, CS-DIV, JS-DIV, α-DIV, C-DIV and the proposed CCS-DIV are discussed. C-DIV, α-DIV and the proposed CCS-DIV with $\alpha = 1, \alpha = 0$ and $\alpha = -1$ are evaluated. Without loss of generality, a simple case has been chosen to elucidate the point. Two binomial variables $\{y_1, y_2\}$ in the presence of the binary events $\{A, B\}$ have been considered as in [7], [13]. The joint probabilities of $p_{y_1,y_2}(A,A), p_{y_1,y_2}(A,B), p_{y_1,y_2}(B,A)$ and $p_{y_1,y_2}(B,B)$ and the marginal probabilities $p_{y_1}(A), p_{y_1}(B), p_{y_2}(A)$ and $p_{y_2}(B)$ are identified. Different divergence methods are tested by fixing the marginal probabilities $p_{y_1}(A) = 0.7, p_{y_1}(B) = 0.3, p_{y_2}(A) = 0.5$ and $p_{y_2}(B) = 0.5$, and setting the joint probabilities of $p_{y_1,y_2}(A,A)$ and $p_{y_1,y_2}(B,A)$ free in intervals (0, 0.7) and (0, 0.3), respectively. Fig. 2 shows the different divergence measures versus the joint probability $p_{y_1,y_2}(A,A)$. All the divergence measures reach to the same minimum at $p_{y_1,y_2}(A,A) = 0.35$ which means that the two random values are independent. Fig. 3 shows the CCS-DIV and α-DIV at different values of α which controls the slope of curves, respectively. Among these measures the steepest curve is obtained by CCS-DIV at $\alpha = -1$. Nonetheless, the CCS-DIV is comparatively sensitive to the probability model and obtains the minimum divergence effectively. However, the CCS-DIV should be a good choice as a contrast function for devising the ICA algorithm. Since, the probability model is closely related to the demixing matrix in the ICA algorithm.

III. CONVEX CAUCHY-SCHWARZ DIVERGENCE INDEPENDENT COMPONENT ANALYSIS (CCS-ICA)

Without loss of generality, we develop the ICA algorithm by using the CCS-DIV as a contrast function. Let us consider a simple system that is described by the vector-matrix form

$$\mathbf{x} = \mathbf{H}\mathbf{s} + \mathbf{v} \quad (9)$$

where $\mathbf{x} = [x_1, \ldots, x_M]^T$ is a mixture observation vector, $\mathbf{s} = [s_1, \ldots, s_M]^T$ is a source signal vector, $\mathbf{v} = [v_1, \ldots, v_M]^T$ is an additive (Gaussian) noise vector, and $\mathbf{H}$ is an unknown full rank M × M mixing matrix, where M is the number of source signals. Let $\mathbf{W}$ be an M × M parameter matrix. To obtain a good estimate, say, $\mathbf{y} = \mathbf{W}\mathbf{x}$ of the source signals $\mathbf{s}$, the contrast function CCS-DIV should be minimized with respect to the demixing filter matrix $\mathbf{W}$. Thus, the components of $\mathbf{y}$ become least dependent when this demixing matrix $\mathbf{W}$ becomes a rescaled permutation of $\mathbf{H}^{-1}$. Following a standard ICA procedure, the estimated source $\mathbf{y}$ can be carried out in two steps: 1) the original data $\mathbf{x}$ should be preprocessed by removing the mean $\{E[\mathbf{x}] = 0\}$ and by a whitening matrix $\{\mathbf{V} = \mathbf{\Lambda}^{-1/2}\mathbf{E}^T\}$, where the matrix $\mathbf{E}$ represents the eigenvectors and the (diagonal) matrix $\mathbf{\Lambda}$ represents the eigenvalues of the autocorrelation of the observations, namely, $\{\mathbf{R}_{\mathbf{xx}} = E[\mathbf{xx}^T]\}$. Consequently, the whitened data vector $\{\mathbf{x}_t = \mathbf{V}\mathbf{x}\}$ would have its covariance equal to the identity matrix, i.e., $\{\mathbf{R}_{\mathbf{x}_t\mathbf{x}_t} = \mathbf{I}_K\}$. The demixing matrix can be iteratively computed by, e.g., the (stochastic) gradient descent algorithm [2]:

$$\mathbf{W}(k+1) = \mathbf{W}(k) - \gamma \frac{\partial D_{CCS}(X, \mathbf{W}(k))}{\partial \mathbf{W}(k)} \quad (10)$$

where $k$ represents the iteration index and $\gamma$ is a step size or a learning rate. Therefore, the updated term in the gradient descent is composed of the differentials of the CCS-DIV with respect to each element $w_{ml}$ of the M × M demixing matrix $\mathbf{W}$. The differentials $\frac{\partial D_{CCS}(X, \mathbf{W}(k))}{\partial w_{ml}(k)}$, $1 \leq m, l \leq M$ are calculated using a probability model and CCS-DIV measures as in [3], [13] and [14]. The update procedure (10) will stop when the absolute increment of the CCS-DIV measure meets a predefined threshold value. During iterations, one should make the normalization step $\mathbf{w}_m = \mathbf{w}_m / \|\mathbf{w}_m\|$ for each row of **W**, where $\|.\|$ denotes a norm. Please refer to *Algorithm 1* for the details about the algorithm based on gradient descent.

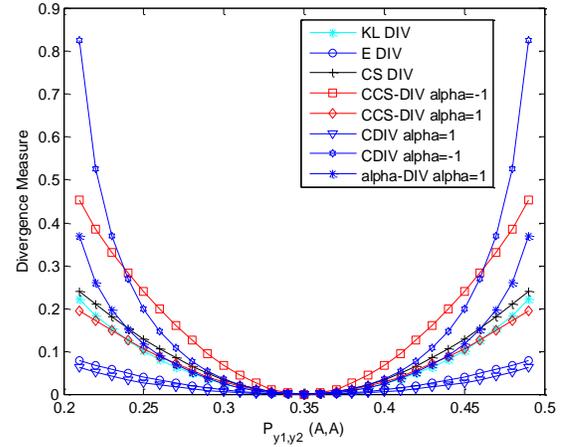

Fig. 2 Different divergence measures versus the joint probability $P_{y_1,y_2}(A,A)$

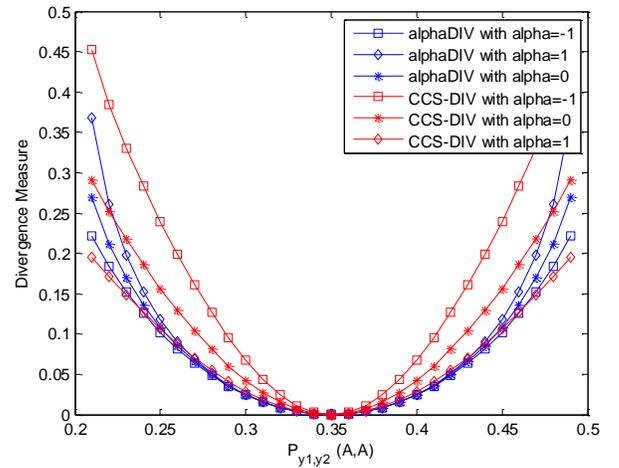

Fig. 3 CCS-DIV and α-DIV versus the joint probability $P_{y_1,y_2}(A,A)$

In deriving the CCS–ICA algorithm, based on the proposed CCS-DIV measure $D_{CCS}(P_J, Q_M, \alpha)$, usually, vector $P_J$ corresponds to the probability of the observed data $\left(p(\mathbf{y}_t) = p(\mathbf{Wx}_t) = \frac{p(\mathbf{x}_t)}{|\det(\mathbf{W})|}\right)$ and vector $Q_M$ corresponds to the probability of the estimated or expected data ($\prod_1^M p(y_{mt}) = \prod_1^M p(\mathbf{w}_m\mathbf{x}_t)$). Here, the CCS–ICA algorithm is detailed as follows. Let the demixed signals $\mathbf{y}_t = \mathbf{Wx}_t$ with its *mth* component denoted as $y_{mt} = \mathbf{w}_m\mathbf{x}_t$. Then, $P_J = p(\mathbf{y}_t) = p(\mathbf{Wx}_t)$ and $Q_M = \prod_1^M p(y_{mt}) = \prod_1^M p(\mathbf{w}_m\mathbf{x}_t)$. Thus, the CCS-DIV as the contrast function, with the built-in convexity parameter α, is

$$D_{CCS}(P_J, Q_M, \alpha)$$
$$= \log \frac{\iint f^2(P_J) dy_1 \dots dy_M \cdot \iint f^2(Q_M) dy_1 \dots dy_M}{[\iint f(P_J) \cdot f(Q_M) dy_1 \dots dy_M]^2}$$
$$= \log \frac{\iint f^2(p(\mathbf{Wx}_t)) dy_1 \dots dy_M \cdot \iint f^2(\prod_1^M p(y_{mt})) dy_1 \dots dy_M}{[\iint f(p(\mathbf{Wx}_t)) \cdot f(\prod_1^M p(y_{mt})) dy_1 \dots dy_M]^2}$$
(11)

For any convex function, we use the Lebesgue measure to approximate the integral with respect to the joint distribution of $y_t = \{y_1, y_2, \dots, y_M\}$. The contrast function thus becomes

$$D_{CCS}(P_J, Q_M, \alpha) = \log \frac{\sum_1^T f^2(p(\mathbf{Wx}_t)) \cdot \sum_1^T f^2(\prod_1^M p(y_{mt}))}{[\sum_1^T f(p(\mathbf{Wx}_t)) \cdot f(\prod_1^M p(y_{mt}))]^2}$$
$$= \log \frac{\sum_1^T f^2(p(\mathbf{Wx}_t)) \cdot \sum_1^T f^2(\prod_1^M (p(w_{mt}\mathbf{x}_t)))}{[\sum_1^T f(p(\mathbf{Wx}_t)) \cdot f(\prod_1^M (p(w_{mt}\mathbf{x}_t)))]^2}$$
(12)

The adaptive CCS–ICA algorithms are carried out by using the derivatives of the proposed divergence, i.e., $\left(\partial D_{CCS}(P_J, Q_M, \alpha) / \partial w_{ml}\right)$ as derived in Appendix A. Note that in Appendix A, the derivative of the determinant demixing matrix ($\det(\mathbf{W})$) with respect to the element ($w_{ml}$) equals the cofactor of entry($m,l$) in the calculation of the determinant of **W**, which we denote as $\left(\frac{\partial \det(\mathbf{W})}{\partial w_{ml}} = W_{ml}\right)$. Also the joint distribution of the output is determined by $p(\mathbf{y}_t) = \frac{p(\mathbf{x}_t)}{|\det(\mathbf{W})|}$.

For simplicity, we can write $D_{CCS}(P_J, Q_M, \alpha)$ as a function of three variables.

$$D_{CCS}(P_J, Q_M, \alpha) = \log \frac{V_1 \cdot V_2}{(V_3)^2}$$
(13)

Then,

$$\frac{\partial D_{CCS}(P_J, Q_M, \alpha)}{\partial w_{ml}} = \frac{V_1' V_2 + V_1 V_2' - 2V_1 V_2 V_3'}{V_1 V_2 V_3}$$
(14)

where

$$V_1 = \sum_{t=1}^{T} f^2(P_J), \quad V_1' = \sum_{t=1}^{T} 2f(P_J)f'(P_J)P_J'$$
$$V_2 = \sum_{t=1}^{T} f^2(Q_M), \quad V_2' = \sum_{t=1}^{T} 2f(Q_M)f'(Q_M)Q_M'$$
$$V_3 = \sum_{t=1}^{T} f(P_J) f(Q_M),$$
$$V_3' = \sum_{t=1}^{T} f'(P_J)f(Q_M)P_J' + \sum_{t=1}^{T} f(P_J)f'(Q_M)Q_M'$$
$$P_J = p(\mathbf{Wx}_t) \text{ and } Q_M = \prod_{m=1}^{M} p(\mathbf{w}_m\mathbf{x}_t)$$
$$P_J' = \frac{\partial P_J}{\partial w_{ml}} = -\frac{p(\mathbf{x}_t)}{|\det(\mathbf{W})|^2} \cdot \frac{\partial \det(\mathbf{W})}{\partial w_{ml}} \cdot \text{sign}(\det(\mathbf{W})),$$

where $\frac{\partial \det(\mathbf{W})}{\partial w_{ml}} = W_{ml}$.

$$Q_M' = \frac{\partial Q_M}{\partial w_{ml}} = \left[\prod_{j=m}^{M} p(\mathbf{w}_j\mathbf{x}_t)\right] \frac{\partial p(\mathbf{w}_n\mathbf{x}_t)}{\partial(\mathbf{w}_n\mathbf{x}_t)} \cdot x_l.$$

where $x_l$ denotes the *lth* entry of $\mathbf{x}_t$.

In general, the estimation accuracy of a demixing matrix in the ICA algorithm is limited by the lack of knowledge of the accurate source probability densities. However, non-parametric density estimate is used in [1], [7], [15], [29 – 32] by applying the effective Parzen window estimation. One of the attributes of the Parzen window is that it must integrate to one. Thus, it is typical to be a pdf itself, e.g., a Gaussian Parzen window, non-Gaussian or other window functions. Furthermore, it exhibits a distribution shape that is data-driven and is flexibly formed based on its chosen Kernel functions. Thus, one can estimate the density function $p(y)$ of the process generating the *M*-dimensional sample $\mathbf{y}_1, \mathbf{y}_2 \dots \mathbf{y}_M$ due to the Parzen Window estimator. For all these reasons, a non-parametric CCS–ICA algorithm is also presented by minimizing the CCS-DIV to generate the demixed signals $\mathbf{y} = [y_1, y_2, \dots, y_M]^T$. The demixed signals are described by the following univariate and multivariate distributions [18]:

$$p(y_m) = \frac{1}{Th} \sum_{t=1}^{T} \vartheta\left(\frac{y_m - y_{mt}}{h}\right) \quad (15)$$

$$p(\mathbf{y}) = \frac{1}{Th^M} \sum_{t=1}^{T} \varphi\left(\frac{\mathbf{y} - \mathbf{y}_t}{h}\right) \quad (16)$$

where the univariate Gaussian Kernel is

$$\vartheta(u) = (2\pi)^{-\frac{1}{2}} e^{-\frac{u^2}{2}}$$

and the multivariate Gaussian Kernel is

$$\varphi(\mathbf{u}) = (2\pi)^{-\frac{N}{2}} e^{\frac{-1}{2}\mathbf{u}^T\mathbf{u}}.$$

The Gaussian kernel(s), used in the non-parametric ICA, are smooth functions. We note that the performance of a learning algorithm based on the non-parametric ICA is better than the performance of a learning algorithm based on the

*Algorithm 1: ICA Based on the gradient descent*

**Input:** $(M \times T)$ matrix of realizations $X$, Initial demixing matrix $W = I_M$, Max. number of iterations Itr, Step Size $\gamma$ i.e. $\gamma = 0.3$, $\alpha$, i.e. $\alpha = -0.99999$

**Perform Pre-Whitening**
$$\{X = V * X = \Lambda^{\wedge}(-1/2)\, E^{\wedge}T\ X\},$$

**For loop:** for each $I$ Iteration do
 **For loop:** for each $t = 1, ..., T$
   Evaluate the proposed contrast function and its
   derivative $\left(\partial D_{CCS}(P_J, Q_M, \alpha) / \partial w_{ml}\right)$
 **End For**
 Update demixing matrix $W$
 $$W = W - \gamma \frac{\partial D_{CCS}(P_J, Q_M, \alpha)}{\partial W}$$
 Check Convergence
 $\|\Delta D_c\| \le \epsilon$ i.e. $\epsilon = 10^{-4}$
**End For**

**Output:** Demixing Matrix $W$, estimated signals $y$

parametric ICA. By substituting (15) and (16) with $y_t = Wx_t$ and $y_{mt} = w_m x_t$ into (12), the nonparametric CCS-DIV becomes

$$P_J = p(y_t) = p(Wx_t) = \frac{1}{Th^M} \sum_{t=1}^{T} \varphi\left(\frac{W(x_t - x_i)}{h}\right)$$

Or

$$P_J = p(y_t) = \frac{p(x_t)}{|\det(W)|}$$

$$Q_M = \prod_{1}^{M} p(y_{mt}) = \prod_{1}^{M} p(w_m x_t)$$

$$= \prod_{1}^{M} \frac{1}{Th} \sum_{i=1}^{T} \vartheta\left(\frac{w_m(x_t - x_i)}{h}\right)$$

$$D_{CCS}(P_J, Q_M, \alpha) = \log \frac{\sum_{t=1}^{T} f^2(P_J) \cdot \sum_{t=1}^{T} f^2(Q_M)}{[\sum_{t=1}^{T} f(P_J) \cdot f(Q_M)]^2} \quad (17)$$

However, there are two common methods to minimize this divergence function: one is based on the gradient descent approach and the other is based on an exhaustive search such as the Jacobi method. In this section, we have presented the derivation of the proposed algorithm in *Appendix A* in order to use it in the non-parametric gradient descent ICA algorithm, see *Algorithm 1*.

## IV. SIMULATION RESULTS

Several illustrative simulation results are conducted to compare the performance of different ICA-based algorithms. This illustration provides results that have a diversity of experimental data and conditions.

### A. Sensitivity of CCS-DIV measure

This experiment evaluates the proposed CCS-DIV divergence measure in relation to the sensitivity of the probability model of the discrete variables. Results indicate that the CCS-DIV with α=1 and α=-1 successfully reach the minimum point of the measure. Let us consider the case as in [13], [14], [15], where the mixed signals X=AS, to investigate the sensitivity of CCS-DIV with α=1 and α=-1, respectively. Simulated experiments in [13], [15] were performed for two sources (M=2) and with a demixing matrix W

$$W = \begin{bmatrix} \cos\theta_1 & \sin\theta_1 \\ \cos\theta_2 & \sin\theta_2 \end{bmatrix} \quad (18)$$

where W, in this case, is a parametrized matrix that establishes a polar coordinate rotation. The row vectors in W have unit norms and provide the counterclockwise rotation of $\theta_1$ and $\theta_2$, respectively. The orthogonal rows in W include the orthogonal matrix rotation when when $\theta_2 = \theta_1 \pm \frac{\pi}{2}$. Notably, the amplitude should not affect the independent sources. By varying $\theta_1$ and $\theta_2$, we get different demixing matrices. However, consider the simple case, i.e., mixtures of signals of two zero mean continuous variables; one variable is of a sub-Gaussian distribution and the other variable is of a super-Gaussian distribution. For the sub-Gaussian distribution, we use the uniform distribution

$$p(s_1) = \begin{cases} \frac{1}{2\tau_1} & s_1 \in (-\tau_1, \tau_1) \\ 0 & \text{Otherwise} \end{cases} \quad (19)$$

and for the super-Gaussian distribution, we use the Laplacian distribution

$$p(s_2) = \frac{1}{2\tau_2} \exp\left[-\frac{|s_2|}{\tau_2}\right] \quad (20)$$

In this task, data samples $T = 1000$ are selected and randomly generated by using $\tau_1 = 3$ and $\tau_2 = 1$. Kurtosis for the two signals are $-1.2$, and $2.99$, respectively, and they are evaluated using $\text{Kurt}(s) = E[s^4] / (E[s^2])^2 - 3$.

Without loss of generality, we take the mixing matrix as the $2 \times 2$ identity matrix, thus, $x_1 = s_1$ and $x_2 = s_2$ [5], [15]. The normalized divergence measures of the demixing signals and their sensitivity to the variation of the demixing matrix is shown in fig. 4. As shown in fig. 4, the variations of the demixing matrix are represented by the polar systems θ1 and θ2. A wide variety of demixing matrices are considered by taking the interval of angles $\{\theta_1 \text{ and } \theta_2\}$ from 0 to $\pi$. Furthermore, fig. 4 evaluates the CCS-DIV along with E-DIV, KL-DIV, and C-DIV with $\alpha = -1$. The minimum (i.e., close to zero) divergence is achieved at the same conditions $\left[\left\{\theta_1 = 0, \theta_2 = \frac{\pi}{2}\right\}, \left\{\theta_1 = \frac{\pi}{2}, \theta_2 = 0\right\}, \left\{\theta_1 = \frac{\pi}{2}, \theta_2 = \pi\right\} \text{ and } \left\{\theta_1 = \pi, \theta_2 = \frac{\pi}{2}\right\}\right]$ as is clearly seen in fig. 4. In addition, one can observe that the CS-DIV does not exhibit a good curvature form in contrast to CCS-DIV from the graphs in Fig. 4. However, the values of CCS-DIV with $\alpha = 1$ are low and flat within the range of $\theta_1$ and $\theta_2$ between 0.5 and 2.5. This performance is similar to other divergence measures as in [13], [15]. Contrarily, the values of CCS-DIV with $\alpha = -1$ enable a relatively increased curvature form in the same range. Thus, the CCS-DIV with $\alpha = -1$ would result in the steepest descent to the minimum point of the CCS-DIV measure.

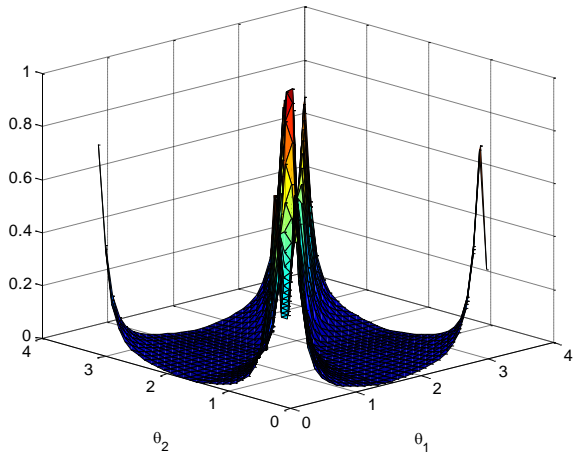

(a) CCS-DIV with α=1

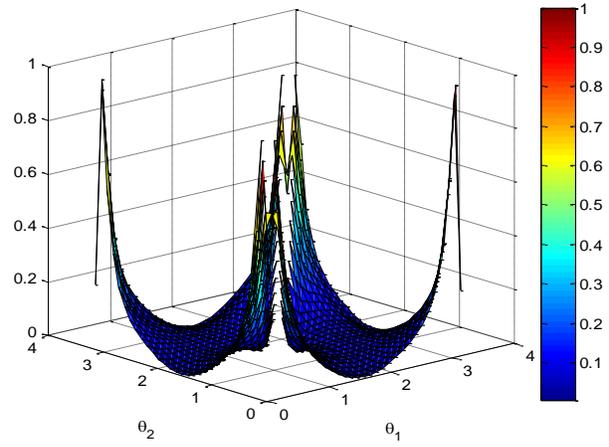

(b) CCS-DIV with α=-1

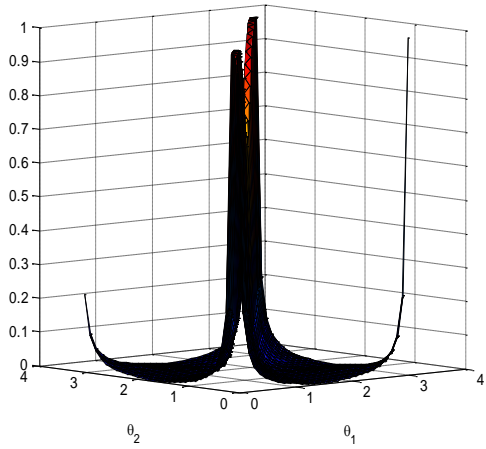

(c) KL-DIV

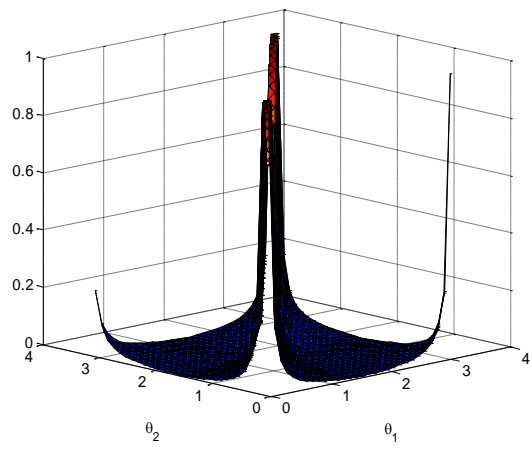

(d) E-DIV

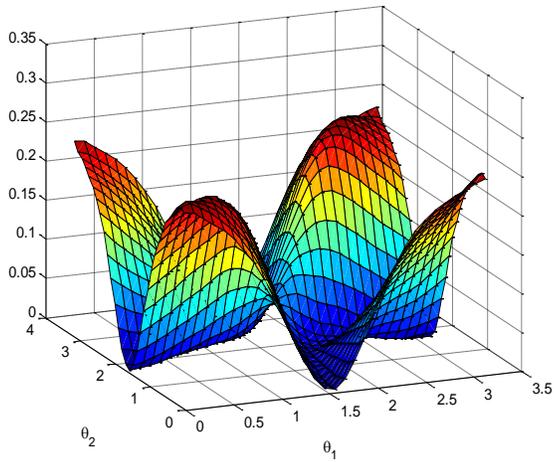

(e) CS-DIV

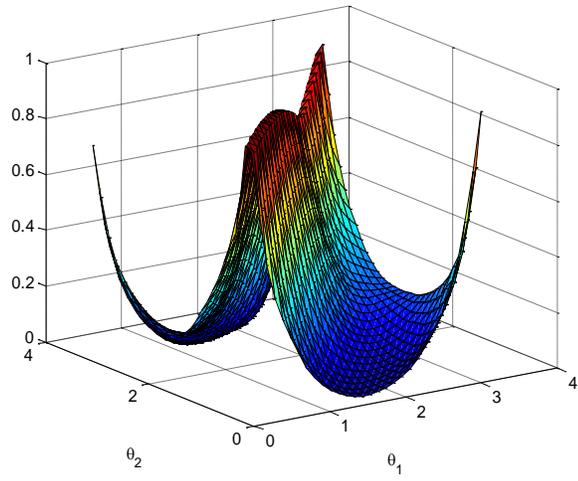

(f) C-DIV with α=-1

Fig. 4. Comparison of (a) CCS-DIV with α = 1, (b) CCS-DIV with α = -1, (c) KL-DIV, (d) E-DIV, (e) CS-DIV and (f) C-DIV with α = -1 of demixed signals as a function of the demixing parameters $\theta_1$ and $\theta_2$.

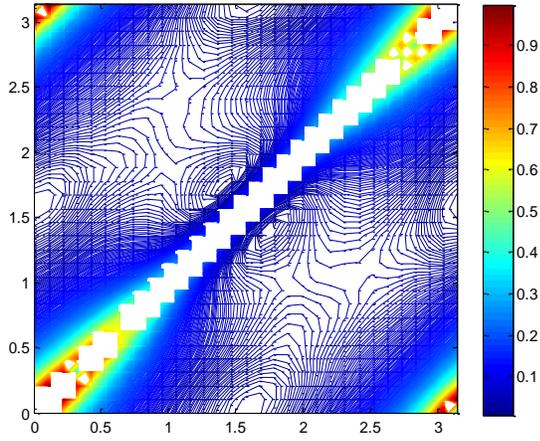

Fig. 5. The contour of the CCS-DIV with α = -1 1 of demixed signals as a function of the demixing parameters $\theta_1$ and $\theta_2$.

Observe that the CCS-DIV with $\alpha = 1$ has a flat curve with respect to the minima in $\theta_1$ and $\theta_2$. For other α values, the CCS-DIV, as a contrast function, can produce large detrimental steps of the demixing matrix towards convergence to a successful demixing solution, as in fig. 5 depicting the contour of CCS-DIV with $\alpha = -1$. It explicitly depicts four locally convex solution minima

### B. Performance evaluation of the proposed CCS-ICA algorithms versus existing ICA-based algorithms

In this section, Monte Carlo Simulations are carried out. It is assumed that the number of sources is equal to the number of observations "sensors". All algorithms have used the same whitening method. The experiments have been carried out using the MATLAB software on an Intel Core i5 CPU 2.4-GHz processor and 4G MB RAM.

First, we compare the performance and convergence speed of the gradient descent ICA algorithms based on the CCS-DIV, CS-DIV, E-DIV, KL-DIV, and C-DIV with $\alpha = 1$ and $\alpha = -1$. In all tasks, the standard gradient descent method is used to devise the parameterized and non-parameterized ICA algorithms based on CCS-DIV with $\gamma = 0.7$ and $\gamma = 0.3$ for α=1 and α=-1 cases, respectively, CS-DIV with $\gamma = 0.3$, E-DIV with γ=0.06, KL-DIV with $\gamma = 0.17$ as in [14], and C-DIV with γ=0.008 and γ=0.1 for the α=-1 and α=1 cases, respectively as in [13]. During the comparison, we use a bandwidth as a function of sample size, namely, $h = 1.06T^{\frac{-1}{5}}$ as in [13-15]. To study the parametric scenario for the ICA algorithms, we use mixed signals that consist of two signal sources with a mixing matrix $A = [[0.5\ 0.6]^T [0.3\ 0.4]^T]$, which has a determinant $det(A) = 0.02$. One of the signal sources has a uniform distribution (sub-Gaussian) and the other has a Laplacian distribution (with kurtosis values -1.2109 and 3.0839, respectively). T = 1000 sampled data are taken using a learning rate γ=0.3 and for 250 iterations. The gradient descent ICA algorithms based on the CCS-DIV, CS-DIV, E-DIV, KL-DIV, and C-DIV with α=1 and α=-1, respectively, are implemented to recover the estimated source signals. The initial demixed matrix W is taken as an identity matrix. Fig. 6 shows the demixed signals resulting from the application of the various ICA-based algorithms. Clearly, the parameterized CCS–ICA algorithm outperforms all other ICA algorithms in this scenario with signal to interference ratio (SIR) of 41.9 dB and 32 dB, respectively. Additionally, Fig. 7 shows the "learning curves" of the parameterized CCS–ICA algorithm with α=1 and α=-1 when compared to the other ICA algorithms, as it graphs the DIV measures versus the iterations (in epochs). As shown in Fig. 7, the speed convergence of the CCS–ICA algorithm is comparable to the C-ICA and KL-ICA algorithms.

### C. Experiments on Speech and Music Signals

Two experiments are presented in this section to evaluate the CCS–ICA algorithm. Both experiments are carried out involving speech and music signals under different conditions. The source signals are two speech signals of different male speakers and a music signal. The first experiment is to separate three source signals from their mixtures given by $X = AS$ where the 3 x 3 mixing matrix

$A = [[0.8\ 0.3\ -0.3]^T [0.2\ -0.8\ 0.7]^T [0.3\ 0.2\ 0.3]^T]$.

The three speech signals are sampled from the ICA '99 conference BSS test sets at http://sound.media.mit.edu/ica-bench/ [13], [15] with an 8 kHz sampling rate. The non-parametrized CCS–ICA algorithms (as well as the other algorithms) with $\alpha = 1$ and $\alpha = -1$ are applied to this task. The resulting waveforms are acquired and the signal to interference ratio (SIR) of each estimated source is calculated. We use the following to calculate the SIR:

Given the source signals $S = \{s_1, s_2, \dots s_M\}$ and demixed signals $Y = \{y_1, y_2, \dots y_M\}$, the SIR in decibels is calculated by

$$\text{SIR (dB)} = 10 \log \frac{\sum_{t=1}^{M} \|s_t\|^2}{\sum_{t=1}^{M} \|y_t - s_t\|^2} \quad (21)$$

The summary results are depicted in Fig. 8. In addition, Fig. 8 shows the SIRs for the other algorithms, namely, JADE[3], Fast ICA[4], Robust ICA[5], KL-ICA and C-ICA with $\alpha = 1$ and $\alpha = -1$. As shown in Fig. 8, the proposed CCS–ICA algorithm achieves significant improvements in terms of SIRs. Also, the proposed algorithm has consistency and obtained the best performance among the host of algorithms.

Moreover, a second experiment is conducted to examine the comparative performance in the presence of additive noise. We now consider the model $\mathbf{x} = A\mathbf{s} + \mathbf{v}$ that contains the same source signals with additive noise and with a different mixing matrix

$A = [[0.8\ 0.3\ -0.3]^T [0.2\ -0.8\ 0.7]^T [0.3\ 0.2\ 0.3]^T]$

The noise $\mathbf{v}$ is an M x T vector with zero mean and $\sigma^2 I$ covariance matrix. In addition, it is independent from the source signals. Fig. 9 shows the separated source signals in the noisy BSS model with SNR = 20 dB. In comparisons, fig. 10 presents the SNRs of all the other algorithms. Clearly, the proposed algorithm has the best performance when compared to others even though its performance decreased in the noisy BSS model. Notably, the SNRs of JADE, Fast ICA and Robust ICA were very low as they rely on the criterion of non-Gaussianity, which is unreliable in the Gaussian-noise environment. In contrast, C-ICA, KL-ICA, and the proposed algorithm, which are based on different mutual information measures, achieved reasonable results. We note that one can also conduct and use the CCS-DIV to recover the source signals from the convolutive mixtures in the frequency domain as in [1], [20]. For brevity, Readers can get more results of non-parametric of CCS-ICA algorithm at http://www.egr.msu.edu/bsr/ .

---
[3] http://www.tsi.enst.fr/icacentral/algos.html
[4] http://www.cis.hut.fi/projects/ica/fastica/code/dlcode.html
[5] http://www.i3s.unice.fr/~zarzoso/robustica.html

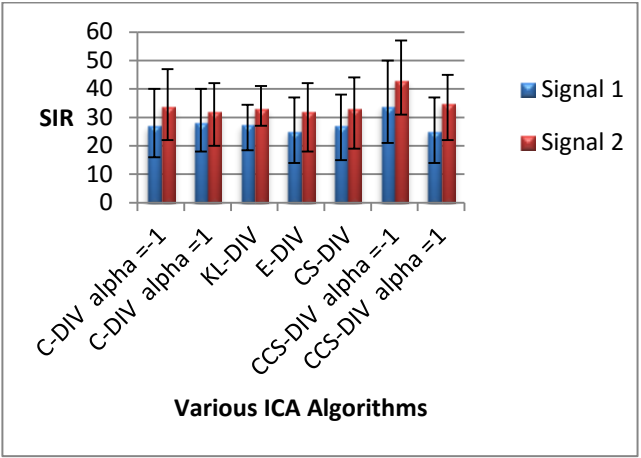

Fig. 6. Comparison of SIRs (dB) of demixed signals by using different ICA algorithms in parametric BSS task.

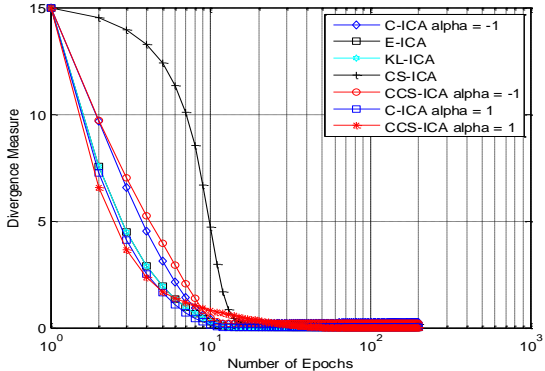

Fig. 7. Comparison of learning curves of C-ICA, E-ICA, KL-ICA, and CCS-ICA with α=1, and α=-1 in a two-source BSS task.

## V. CONCLUSION

A new divergence measure has been presented based on integrating a convex function into Cauchy-Schwarz inequality. This divergence measure has been used as a contrast function in order to build a new ICA algorithm to solve the BSS (Blind Source Separation) problem. The CCS-DIV attains steeper decent to the solution point. Sample experiments and examples are carried out to show the performance of the proposed divergence. This paper has developed the nonparametric CCS-ICA approach using the Parzen window density estimate. The proposed CCS-ICA achieved the highest SIR in separation of tested speech and music signals relative to other ICA algorithms.

## APPENDIX A

### THE CONVEX CAUCHY–SCHWARZ DIVERGENCE AND ITS DERIVATIVE

Assume the demixed signals are given by $\mathbf{y_t} = \mathbf{Wx_t}$ where the $mth$ component is $y_{mt} = \mathbf{w_m x_t}$. Now, express the CCS-DIV as a contrast function with a convexity parameter α (in f) as follows:

$$D_{CCS}(\mathbf{X}, \mathbf{W}, \alpha)$$
$$= \log \frac{\iint f^2(p(\mathbf{y_t})) dy_1 \dots dy_M \cdot \iint f^2(\prod_1^M p(y_{mt})) dy_1 \dots dy_M}{[\iint f(p(\mathbf{y_t})) \cdot f(\prod_1^M p(y_{mt})) dy_1 \dots dy_M]^2}$$

By using the Lebesgue measure to approximate the integral with respect to the joint distribution of $\mathbf{y_t} = \{y_1, y_2, \dots, y_M\}$, the contrast function becomes

$$D_{CCS}(\mathbf{X}, \mathbf{W}, \alpha) = \log \frac{\sum_1^T f^2(p(\mathbf{Wx_t})) \cdot \sum_1^T f^2(\prod_1^M (p(\mathbf{w_m x_t})))}{[\sum_1^T f(p(\mathbf{Wx_t})) \cdot f(\prod_1^M (p(\mathbf{w_m x_t})))]^2}$$

For simplicity, let us assume

$$V_1 = \sum_{t=1}^T f^2(\mathbf{y_t}), \quad V_1' = \sum_{t=1}^T 2f(\mathbf{y_t}) f'(\mathbf{y_t}) \mathbf{y_t'}$$

$$V_2 = \sum_{t=1}^T f^2(y_{mt}), \quad V_2' = \sum_{t=1}^T 2f(y_{mt}) f'(y_{mt}) y_{mt}'$$

$$V_3 = \sum_{t=1}^T f(\mathbf{y_t}) f(y_{mt}),$$

$$V_3' = \sum_{t=1}^T f'(\mathbf{y_t}) f(y_{mt}) \mathbf{y_t'} + \sum_{t=1}^T f(\mathbf{y_t}) f'(y_{mt}) y_{mt}'$$

and the convex function is (e.g.)

$$f(t) = \frac{4}{1-\alpha^2} \left[ \frac{1-\alpha}{2} + \frac{1+\alpha}{2} t - t^{\frac{1+\alpha}{2}} \right]$$

$$f'(t) = \frac{2}{1-\alpha} \left[ 1 - t^{\alpha-1/2} \right]$$

then,

$$\mathbf{y_t} = p(\mathbf{Wx_t}) \text{ and } y_{mt} = \prod_{m=1}^M p(\mathbf{w_m x_t})$$

$$\mathbf{y_t'} = \frac{\partial \mathbf{y_t}}{\partial w_{ml}} = -\frac{p(\mathbf{x_t})}{|\det(\mathbf{W})|^2} \cdot \frac{\partial \det(\mathbf{W})}{\partial w_{ml}} \cdot \text{sign}(\det(\mathbf{W})),$$

where $\frac{\partial \det(\mathbf{W})}{\partial w_{ml}} = W_{ml}$;

$$y_{mt}' = \frac{\partial y_{mt}}{\partial w_{ml}} = \left[ \prod_{j \neq m}^M p(\mathbf{w_j x_t}) \right] \frac{\partial p(\mathbf{w_m x_t})}{\partial(\mathbf{w_m x_t})} \cdot x_l.$$

where $x_l$ denotes the $l^{th}$ entry of $\mathbf{x_t}$.

Thus, we re-write the CCS-DIV as

$$D_{CCS}(\mathbf{X}, \mathbf{W}, \alpha) = \log \frac{V_1 \cdot V_2}{[V_3]^2} = \log V_1 + \log V_2 - 2 \log V_3$$

and its derivative becomes

$$\frac{\partial D_{CCS}(\mathbf{X}, \mathbf{W}, \alpha)}{\partial w_{ml}} = \frac{V_1'}{V_1} + \frac{V_2'}{V_2} - 2 * \frac{V_3'}{V_3}$$

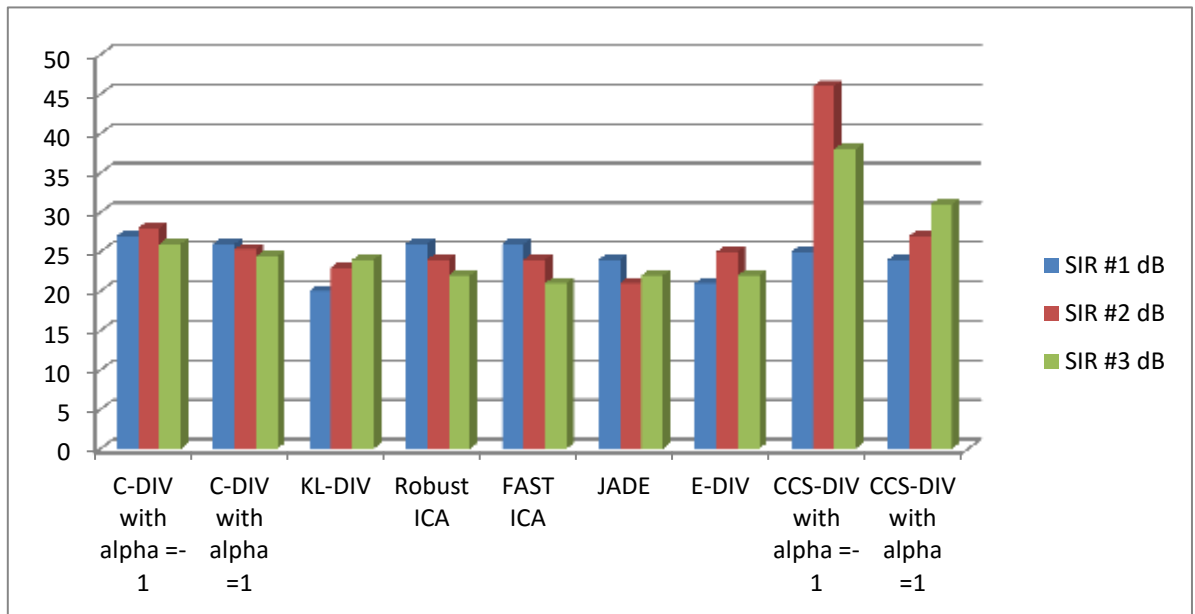

Fig. 8. Comparison of SIRs (dB) of demixed two speeches and music signals by using different ICA algorithms in instantaneous BSS tasks.

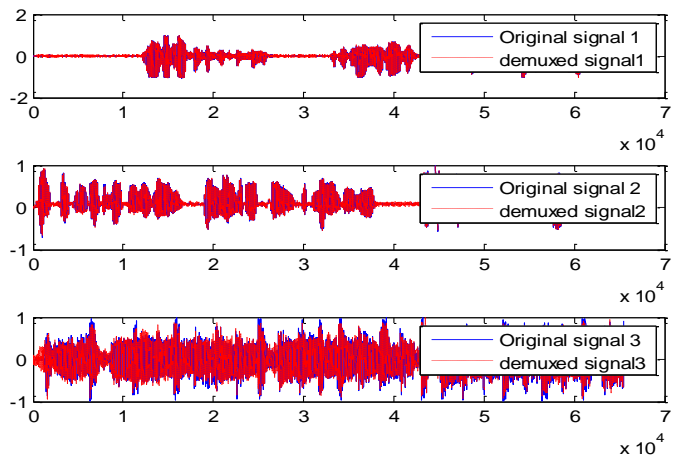

Fig. 9. . the original signals and de-mixed signals by using CCS-ICA algorithm in instantaneous BSS tasks with additive Gaussian noise.

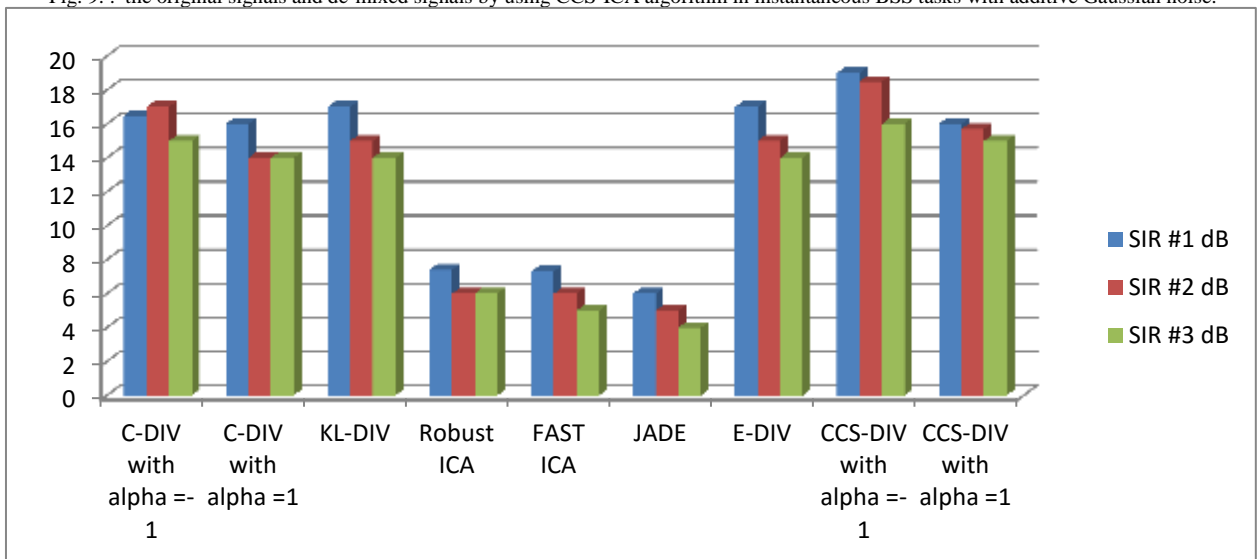

Fig. 10. Comparison of SIRs (dB) of demixed two speeches and music signals by using different ICA algorithms in instantaneous BSS tasks with additive Gaussian noise.